\documentclass{aastex63}
\usepackage{gensymb}
\usepackage{booktabs}
\usepackage[T1]{fontenc}
\usepackage[utf8]{inputenc}
\usepackage{babel}
\usepackage{threeparttable}  
\usepackage{makecell}
\usepackage{multirow}
\usepackage{amsmath}
\usepackage{diagbox}
\usepackage{stackengine}
\usepackage[figuresright]{rotating}
\newcommand\xrowht[2][0]{\addstackgap[.5\dimexpr#2\relax]{\vphantom{#1}}}

\submitjournal{APJS}

\shorttitle{A catalog of LAMOST variable sources based on time-domain photometry of ZTF}
\shortauthors{Xu et al.}

\begin{document}

\title{A Catalog of LAMOST Variable Sources Based on Time-domain Photometry of ZTF
    \footnote{Released on ***, ***, ***}}

\correspondingauthor{Feng Wang}
\email{fengwang@gzhu.edu.cn}

\author[0000-0002-9997-9524]{Tingting Xu}
\affil{Center For Astrophysics, Guangzhou University,
Guangzhou, Guangdong, China, 510006}
\affil{Great Bay Center, National Astronomical Data Center, Guangzhou, Guangdong, China, 510006}
\affil{Astronomy Science and Technology Research Laboratory of Department of Education of Guangdong Province, 
Guangzhou, Guangdong, China, 510006}
\affil{Key Laboratory for Astronomical Observation and Technology of Guangzhou,Guangzhou, Guangdong, China, 510006}

\author[0000-0002-1802-6917]{Chao Liu}
\affil{Key Laboratory of Space Astronomy and Technology, National Astronomical Observatories, CAS,
Beijing, China,100101}
\affil{University of Chinese Academy of Sciences,
Beijing, China,100049}

\author[0000-0002-9847-7805]{Feng Wang}
\affil{Center For Astrophysics, Guangzhou University,
Guangzhou, Guangdong, China, 510006}
\affil{Great Bay Center, National Astronomical Data Center, Guangzhou, Guangdong, China, 510006}
\affil{Astronomy Science and Technology Research Laboratory of Department of Education of Guangdong Province, 
Guangzhou, Guangdong, China, 510006}
\affil{Key Laboratory for Astronomical Observation and Technology of Guangzhou,Guangzhou, Guangdong, China, 510006}
\affil{University of Chinese Academy of Sciences,
Beijing, China,100049}

\author[0000-0001-8449-6020]{Weirong Huang}
\affil{Center For Astrophysics, Guangzhou University,
Guangzhou, Guangdong, China, 510006}
\affil{Great Bay Center, National Astronomical Data Center, Guangzhou, Guangdong, China, 510006}
\affil{Astronomy Science and Technology Research Laboratory of Department of Education of Guangdong Province, 
Guangzhou, Guangdong, China, 510006}
\affil{Key Laboratory for Astronomical Observation and Technology of Guangzhou,Guangzhou, Guangdong, China, 510006}

\author[0000-0002-8765-3906]{Hui Deng}
\affil{Center For Astrophysics, Guangzhou University,
Guangzhou, Guangdong, China, 510006}
\affil{Great Bay Center, National Astronomical Data Center, Guangzhou, Guangdong, China, 510006}
\affil{Astronomy Science and Technology Research Laboratory of Department of Education of Guangdong Province, 
Guangzhou, Guangdong, China, 510006}
\affil{Key Laboratory for Astronomical Observation and Technology of Guangzhou,Guangzhou, Guangdong, China, 510006}

\author[0000-0002-7960-9251]{Ying Mei}
\affil{Center For Astrophysics, Guangzhou University,
Guangzhou, Guangdong, China, 510006}
\affil{Great Bay Center, National Astronomical Data Center, Guangzhou, Guangdong, China, 510006}
\affil{Astronomy Science and Technology Research Laboratory of Department of Education of Guangdong Province, 
Guangzhou, Guangdong, China, 510006}
\affil{Key Laboratory for Astronomical Observation and Technology of Guangzhou,Guangzhou, Guangdong, China, 510006}

\author{Zhong Cao}
\affil{Center For Astrophysics, Guangzhou University,
Guangzhou, Guangdong, China, 510006}
\affil{Great Bay Center, National Astronomical Data Center, Guangzhou, Guangdong, China, 510006}
\affil{Astronomy Science and Technology Research Laboratory of Department of Education of Guangdong Province, 
Guangzhou, Guangdong, China, 510006}
\affil{Key Laboratory for Astronomical Observation and Technology of Guangzhou,Guangzhou, Guangdong, China, 510006}


\begin{abstract}

The identification and analysis of different variable sources is a hot issue in astrophysical research.
The Large Sky Area Multi-Object Fiber Spectroscopic Telescope (LAMOST) spectroscopic survey has accumulated massive spectral data but contains no information about variable sources.
Although a few related studies present variable source catalogs for the LAMOST, the studies still have a few deficiencies regarding the type and number of variable sources identified.
In this study, we presented a statistical modeling approach to identify variable source candidates.
We first crossed the Kepler, Sloan Digital Sky Survey (SDSS), and Zwicky Transient Facility (ZTF) catalogs to obtain light curves data of variable and non-variable sources. 
The data are then modeled statistically using commonly used variability parameters, respectively. And then, an optimal variable source identification model is determined using the Receiver Operating Characteristic (ROC) curve and four credible evaluation indices such as precision, accuracy, recall, and F1\_score. Based on this identification model, a catalog of LAMOST variable sources (including 631,769 variable source candidates with a probability greater than 95$\%$ and so on) is obtained. To validate the correctness of the catalog, we performed a two-by-two cross-comparison with the GAIA catalog and other published variable source catalogs. We achieved the correct rate ranging from 50\% to 100\%. Among the 123,756 sources cross-matched, our variable source catalog identifies 85,669 with a correct rate of 69\%, which indicates that the variable source catalog presented in this study is credible.

\end{abstract}
 
                                    

\section{Introduction} \label{sec:intro}

The research of variable sources has been one of the frontier topics in astronomy research and is also the core of many research topics in astrophysics (\cite{EyerMowlavi2008}). For example, eruptive and episodic systems can improve the understanding of accretion, mass loss, and stellar birth(\cite{Crawford1955}). Eclipsing systems constrain exoplanet demographics, mass transfer, binary evolution, and the mass-radius-temperature relation of stars. Pulsating sources are essential for probing stellar structure and stellar evolution theory(\cite{Walkowiczetal2009}). In addition, some eclipsing systems and many of the most common pulsating systems (e.g., RR Lyrae, Cepheids, and Mira variables) are the fundamental means to measure precise distances to clusters, to the local group of galaxies, and to relic streams of disrupted satellites around the Milky Way(\cite{Richardsetal2011},\cite{Kim2003}).

Although variable sources have been studying for several hundred years, the discovery and identification of many variable sources rely heavily on the emergence of new detection tools. With the advent of the Charge-Coupled Device (CCD), the Optical Gravitational Lensing Experiment(OGLE) detected more than 900,000 variables during its 20-year observation(\cite{udalski1992optical},\cite{udalski2015ogle}). The first all-sky variability survey was carried out by the All-Sky Automated Survey (ASAS), which contained the eclipsing binaries and periodic pulsating sources(\cite{pojmanski2005all}). Meanwhile, a series of catalogs of variable sources have been published based on different types of telescope(\cite{drake2014catalina}, \cite{drake2017catalina}, \cite{cioni2011vmc}). Subsequently, the development of time-domain astronomy has further promoted the identification of variable sources. A number of variable source catalogs were published, including the Wide-field Infrared Survey Explorer(WISE) catalog of periodic variable sources(\cite{chenetal2018b}), the variable catalogs of Gaia DR2 (\cite{Clementinietal2019}) and DR3 (\cite{brown2020gaia}), the Zwicky Transient Facility(ZTF) catalog of periodic variable sources(\cite{Chenetal2020}), and the catalog of over ten million variable source candidates based on ZTF DR1(\cite{ofek2020}).

The Large Sky Area Multi-Object Fiber Spectroscopic Telescope(LAMOST) (\cite{Zhaoetal2012},\cite{cui2012}) is a Schmidt telescope with an effective aperture of 3.6 $-$ 4.9m with the field of view of about 5$^{\circ}$, which was designed to collect 4000 spectra in a single exposure (spectral resolution R $\sim$ 1800, limiting magnitude r $\sim$ 19 mag, wavelength coverage 3700 $-$ 9000Å). 
LAMOST survey contains two main topics: the LAMOST ExtraGAlactic Survey (LEGAS) and the LAMOST Experiment for Galactic Understanding and Exploration (LEGUE) survey of the Milky Way stellar structure(\cite{Paunzenetal2020}). At the end of September 2020, the LAMOST DR6 catalog was officially released to the world. It includes the 4901 observation areas and 11.27 million spectra data. 
LAMOST provides valuable information on atmospheric parameters, which are essential parameters for variable source studies and can help make more accurate classifications and improve the classification accuracy of variable sources in combination with light curves. However, there are no variable sources information in the current LAMOST catalog, imposing significant limitations on using LAMOST data for variable sources studies. 

The issue of variable source identification for LAMOST catalogs has been studied in several related studies, and the corresponding catalogs have been published. \cite{Tianetal2020} presented a LAMOST radial velocity variable sources catalog. They first selected the sources observed multiple times in LAMOST and compared the observed radial velocity variation with the simulated radial velocity variation to estimate the probability of the radial velocity variable source. They finally obtained 80,702 radial velocity variable sources with a probability greater than 60\%, and 3138 sources have been classified by crossing the data from other published catalogs. In addition, cross-matching with other catalogs is also a very effective way to identify variable sources for the LAMOST. \cite{ofek2020} obtained 10 million variable source data by standard deviation and period analysis methods, which contains several new short-period (less than 90 minutes) candidates, about 60 new dwarf nova candidates, two candidate eclipsing systems, and so on. They also provided cross-matching information with catalogs such as GAIA, SDSS, and LAMOST.

These works have effectively advanced the identification of variable sources for the LAMOST. However, these works still have some minor limitations. For example, the identification method of \cite{Tianetal2020} is based on the radial velocity. The identification requires a sufficient number of observations of the variable source candidates. This results in a small number of variables and a low identification rate for other variable source types.
The study of \cite{ofek2020} is mainly based on the periodic analysis method, which works well for periodic variable sources but somewhat limits the discovery of other types of variable sources, especially non-periodic or long-period variables.  
In addition, the variable source catalogs presented by \cite{ofek2020} and \cite{Chenetal2020} are limited in the number of sources that can be crossed with the LAMOST catalog. Although the catalog of \cite{ofek2020} has 10 million variable sources, only 216,007 sources have been observed by the LAMOST. Within 216,007 sources, 58\% of the data with a signal-to-noise ratio in g band(SNRg) less than 15, and 66\% of the data with SNRg less than 20. According to the data quality statement published on the LAMOST official website, data with SNR less than 15 generally suffer from significant uncertainties in either radial velocity or stellar parameter estimation.

In this study, we tried a statistical modeling approach for the identification of the LAMOST variable sources. The specific implementation is described in section~\ref{sec:methods} in detail. We further apply these models to the data sets of LAMOST DR6 and ZTF DR2 and get the final catalog of LAMOST variable source candidates in section~\ref{sec:catalog}. In section~\ref{sec:crossmatch}, the catalog of LAMOST variable source candidates is evaluated by cross-identifying with the other variable sources catalogs obtained from different survey projects. And then, we make a discussion for models and results in section~\ref{sec:Discussion}. Finally, we present the conclusion in section~\ref{sec:Conclusion}.

\section{The Identification Approach of Variable Sources Based on Light Curves} \label{sec:methods}

\begin{figure}[h]
    \centering
    \includegraphics[width=12cm]{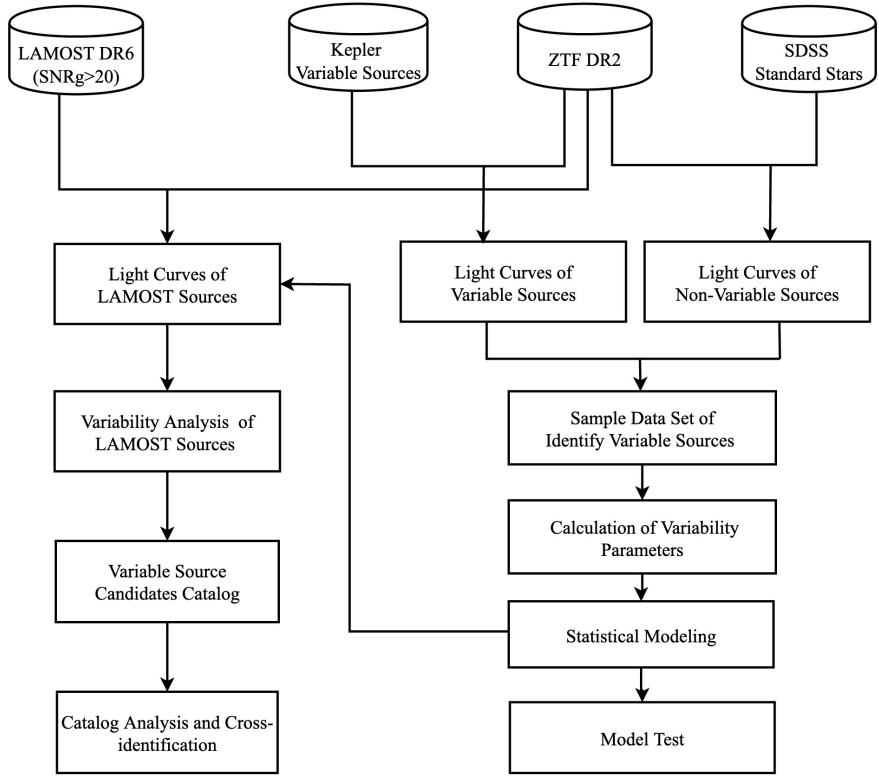} 
    \caption{The data-flow diagram of identifying variable sources for LAMOST sources.}
    \label{fig-dataflow}
\end{figure}

A statistical modeling approach was used in the study. The main idea is to first model the variability parameters by the light curves of variable and non-variable sources. Then the optimal model for the variability parameters is determined by model testing. Finally, the model is used to identify the variable source candidates. A data-flow diagram introducing modeling identification is shown in Fig.~\ref{fig-dataflow}. The entire statistical modeling process is divided into four main parts: searching trustful identification parameters, generating a sample set of light curves, creating statistical modeling for identification, and model testing.

\subsection{Variability Parameters } \label{subsection:parameters}

The variability parameters of light curves are the keys for statistical modeling, ranging from basic statistical properties (e.g., mean, standard deviation, and so on) to more complex time series characteristics (e.g., auto-correlation function). 

According to the characteristics of light curves, we select seven variability parameters, i.e., Std, Iter-std, $C_\nu$, $\kappa$, $\gamma$, MAD, and Amp, from previous studies (\cite{Nun2015},\cite{Nun2017}). These variability parameters had been proven to work well for classifying variable sources through a machine learning approach (\cite{cabraletal2018}, \cite{Coughlin2021}, \cite{vanRoestel2021}). In addition, we introduce three variability parameters (Q, Q1, and Q2). All parameters are described as follows.

(1) Q value(Q):

\begin{equation}
     Q = \frac{\left| {m_{max} - m_{min}} \right|} 
     {\sqrt{{\sigma_{max}^2 +\sigma_{min}^2}}}
\end{equation}
where $m_{max}$ and $m_{min}$ are the maximum and minimum magnitude in the light curves, respectively. $\sigma_{max}$ and $\sigma_{min}$ are their magnitude measurement errors. 

(2)Q1 value(Q1):

Q1 is the variant form of Q. After the maximum and minimum magnitude of the light curves are removed, we re-calculated this parameter by the same calculation method as Q.

(3)Q2 value(Q2):

Q2 is also the variant form of Q. After removing the maximum, minimum, sub-maximum, and sub-minimum magnitudes of the light curves, re-calculated this parameter by the same calculation method as Q.

(4) Standard deviation(Std):

\begin{equation}
     \sigma = \sqrt{{\frac{1}{N - 1}} {\sum \limits_{i}(m_{i} - \overline{m})}}
\end{equation}
where $N$ is the detection times of light curves from the ZTF catalog, m$_{i}$ is the magnitude of each observation in the light curves, and $\overline{m}$ is the mean of magnitude. 

(5) Iterative standard deviation(Iter-std):

After calculating the standard deviation of the light curve, the data other than the median plus or minus twice the standard deviation are removed, and the standard deviation is re-calculated. This process is repeated until the resulting standard deviation converges to a stable value.

(6) Coefficient of variation(C$_{\nu}$):

\begin{equation}
     C_{\nu}= \frac{\sigma}{\overline{m}}
\end{equation}
The $C_{\nu}$ is a simple variability index and is defined as the ratio of the standard deviation to the mean magnitude. If a light curve has substantial variability, the $C_{\nu}$ of this light curve is generally significant.

(7)Small Kurtosis($\kappa$):

Small sample kurtosis of the magnitudes:
\begin{equation}
    \kappa = \frac{N(N+1)}{(N-1)(N-2)(N-3)}
    \sum \limits_{i=1}^{N}(\frac{m_{i}-\hat{m}}{\sigma})^4 
    - \frac{3(N-1)^2}{(N-2)(N-3)}
\end{equation}
Where $\hat{m}$ is the median of magnitude. For a normal distribution, the small Kurtosis should be zero.

(8)Skewness($\gamma$):

The skewness of a source is defined as follow:
\begin{equation}
    \gamma = \frac{N)}{(N-1)(N-2)}
    \sum \limits_{i=1}^{N}(\frac{m_{i}-\hat{m}}{\sigma})^3
\end{equation}
For a normal distribution, it should be equal to zero.

(9)Median absolute deviation(MAD):

The MAD is described as the median discrepancy of the data from the median data:
 \begin{equation}
     MAD = \hat{m} (|m_{i}-\hat{m}|)
 \end{equation}
A normal distribution should have a value of about 0.675. The interquartile ranges of a normal distribution can be used to illustrate this.
 
(10)Amplitude(Amp):

The amplitude is half of the difference between the median of the maximum and minimum 5$ 
\% $ magnitudes. The amplitude of a set of numbers from 0 to 1000 should be 475.5.

\subsection{Data Set Preparation} \label{subsection:sample}

As described in the previous section, the statistical modeling approach seriously depends on data samples from sources. Constructing credible data set from variable and non-variable sources is the key to modeling.

1. Variable source data set

We build a variable source data set based on the variable sources explicitly labeled in the Kepler-related catalogs. The Kepler Space Telescope makes long-term continuous photometric observations in specific sky regions. A series of variable source catalogs based on these photometric data  are published (\cite{Nielsenetal2013}, \cite{Abdul-Masih2016}, \cite{Reinhold2013},\cite{Bowman2016}). The variable sources in the Kepler catalogs are accurately labeled.
We first selected 12,151 rotating variable sources (\cite{Nielsenetal2013}) and 983 pulsating variable sources (\cite{Bowman2016}). In addition, a catalog of 2926 eclipsing binaries was selected from the Kepler website (http://keplerebs.villanova.edu/). A total of 3,752 common sources were obtained by cross-matching these Kepler variable sources with the ZTF catalog, and the light curves data for these sources were obtained from the ZTF DR2.

2. Non-variable source data set

To ensure the correctness of the non-variable sources, we finally select the light curves of the standard stars as the non-variable source dataset. Standard stars are used as benchmarks in astrophysics observations such as photometry and spectral classification, and usually, the magnitude of light curves is constant. Based on the multiple photometric observations of the Sloan Digital Sky Survey(SDSS), \cite{Ivezi2007} provided a standard star catalog that includes 1.01 million non-variable unresolved objects. All sources are located near the equator. A total of 263,670 sources are obtained by cross-matching the SDSS standard stars catalog with ZTF DR2 catalogs. We also collect the light curves of these sources from ZTF DR2 to build the non-variable data set.

\begin{figure}[ht]
    \centering
    \includegraphics[width=16cm,height = 13cm]{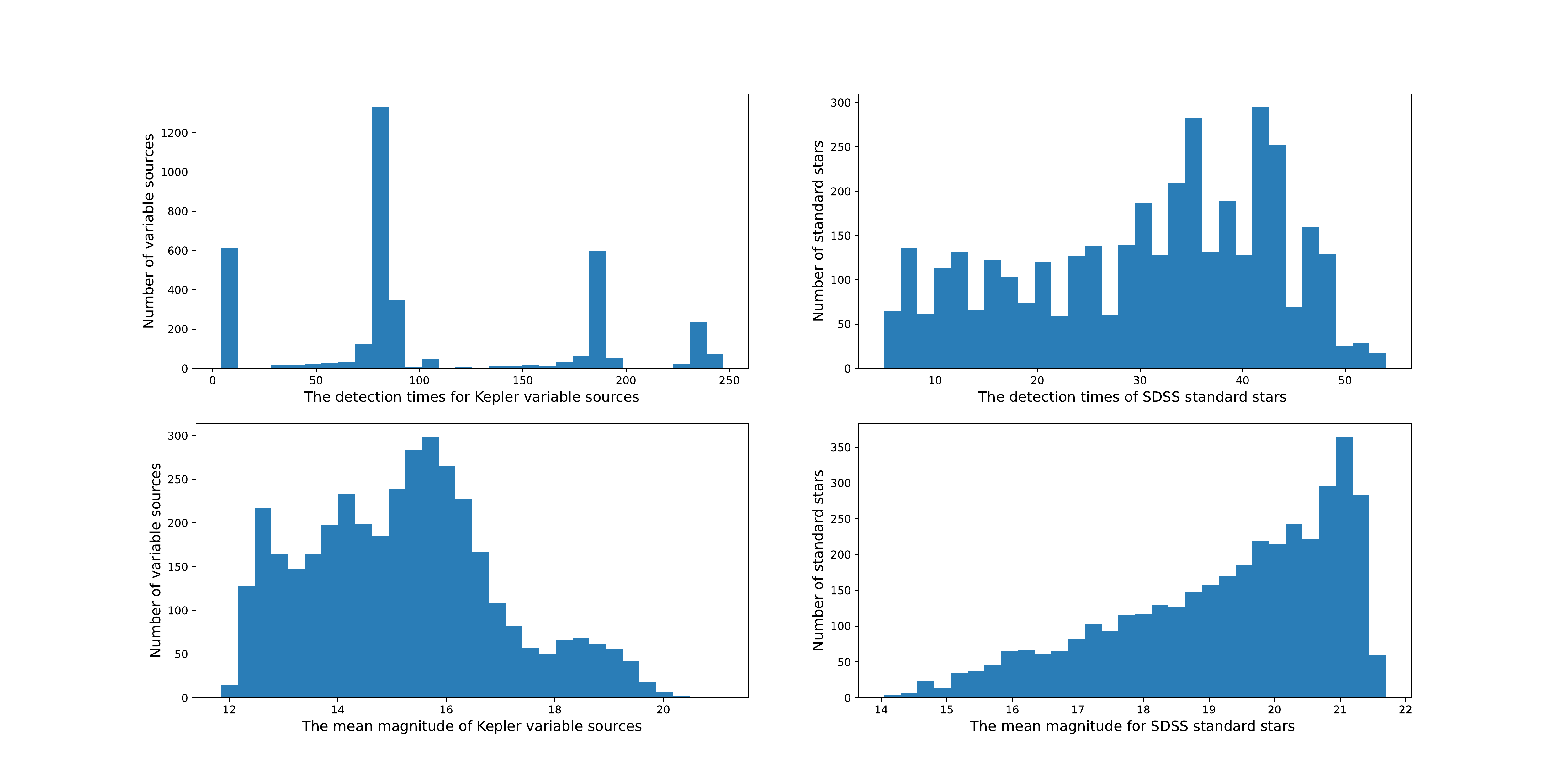} 
    \caption{The statistical histograms of detection times and mean magnitude vs. variable sources/standard stars distributions.}
    \label{sample_Nobs_mag}
\end{figure}

And then, we randomly select SDSS standard stars equal to the number of Kepler variable sources. The total number of the sample data set that we used in the statistical modeling is 7,504. The distribution of detection times and mean magnitude of light curves for these sources is shown in Fig.~\ref{sample_Nobs_mag}. It shows the detection times of the standard stars are significantly less than that of the variable sources, and most of the standard stars are faint.

\begin{figure}[ht]
    \centering
    \includegraphics[width=17cm,height = 8cm]{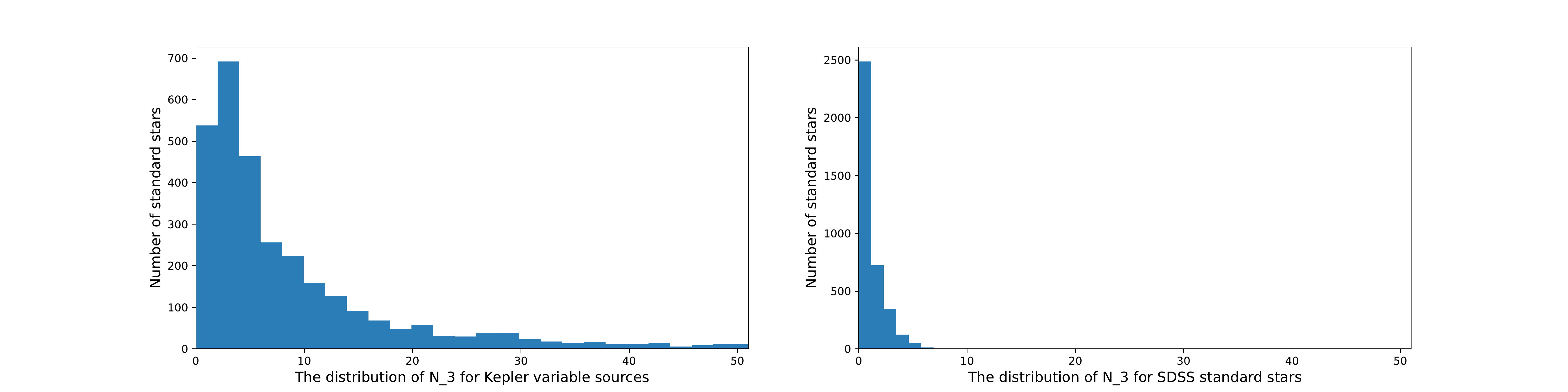} 
    \caption{The statistical histograms of sigma level  vs. variable sources$/$standard stars distributions.}
    \label{sigma_level}
\end{figure}

To further determine the availability of the standard stars, we counted the magnitude error distribution of the selected standard stars (see Fig.~\ref{sigma_level}). From the statistical histograms, the magnitude errors are within the 3 sigma level(N\_3) for basically all the standard stars, which indicates that the light curves of the standard stars are essentially constant.

\begin{figure}[ht]
    \centering
    \includegraphics[width=16cm,height = 13cm]{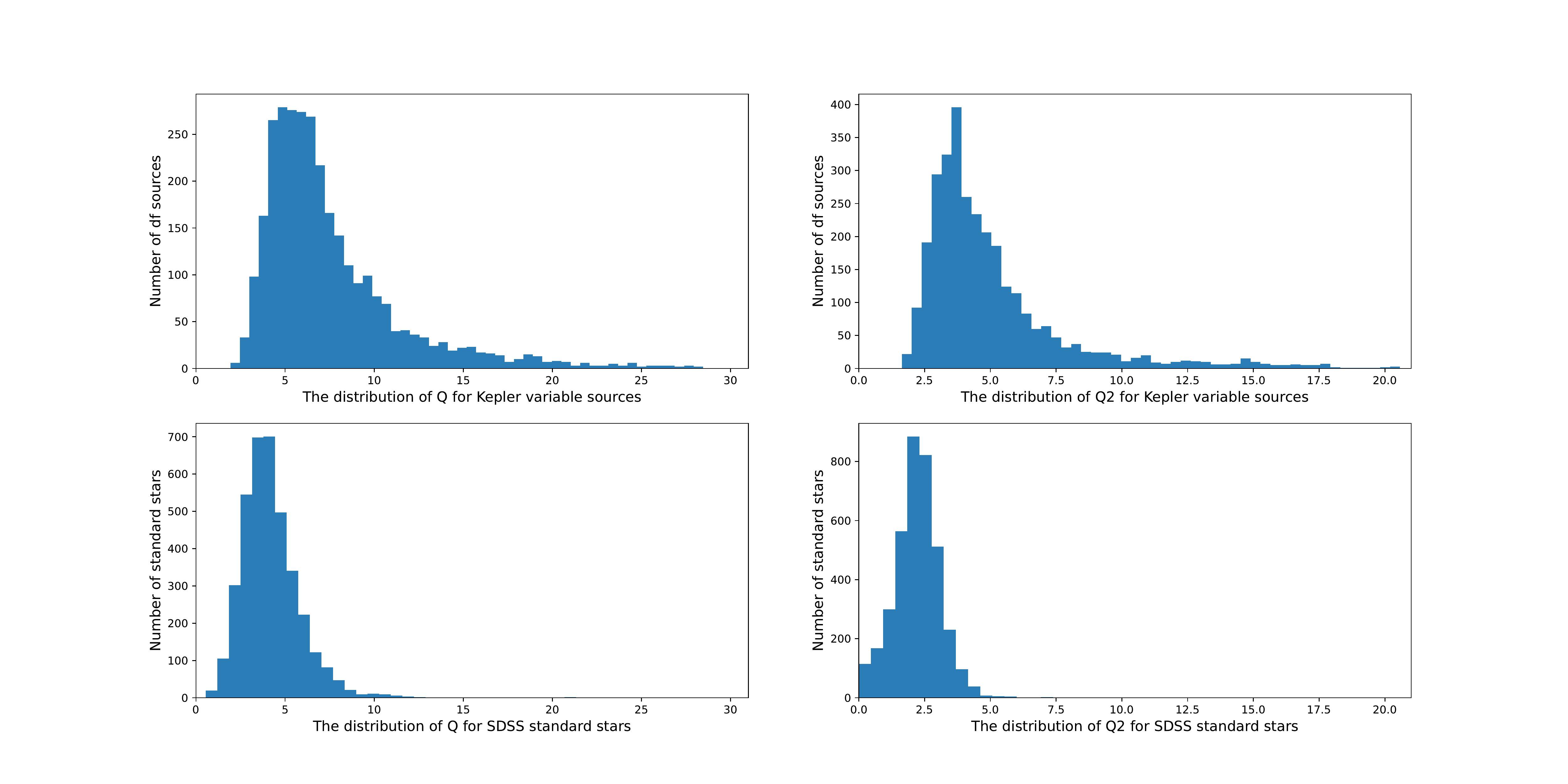} 
    \caption{The statistical histograms of Q and Q2 vs. variable sources$/$standard stars distributions.}
    \label{fig-datadistribution}
\end{figure}

Based on the sample data set, we calculate the values of each of the ten variability parameters. The results show significant differences between the values of the ten parameters calculated by the variable sources and the non-variable sources (see Fig.~\ref{fig-datadistribution}). 

\begin{figure}[ht]
    \centering
    \includegraphics[width=10cm,height = 8cm]{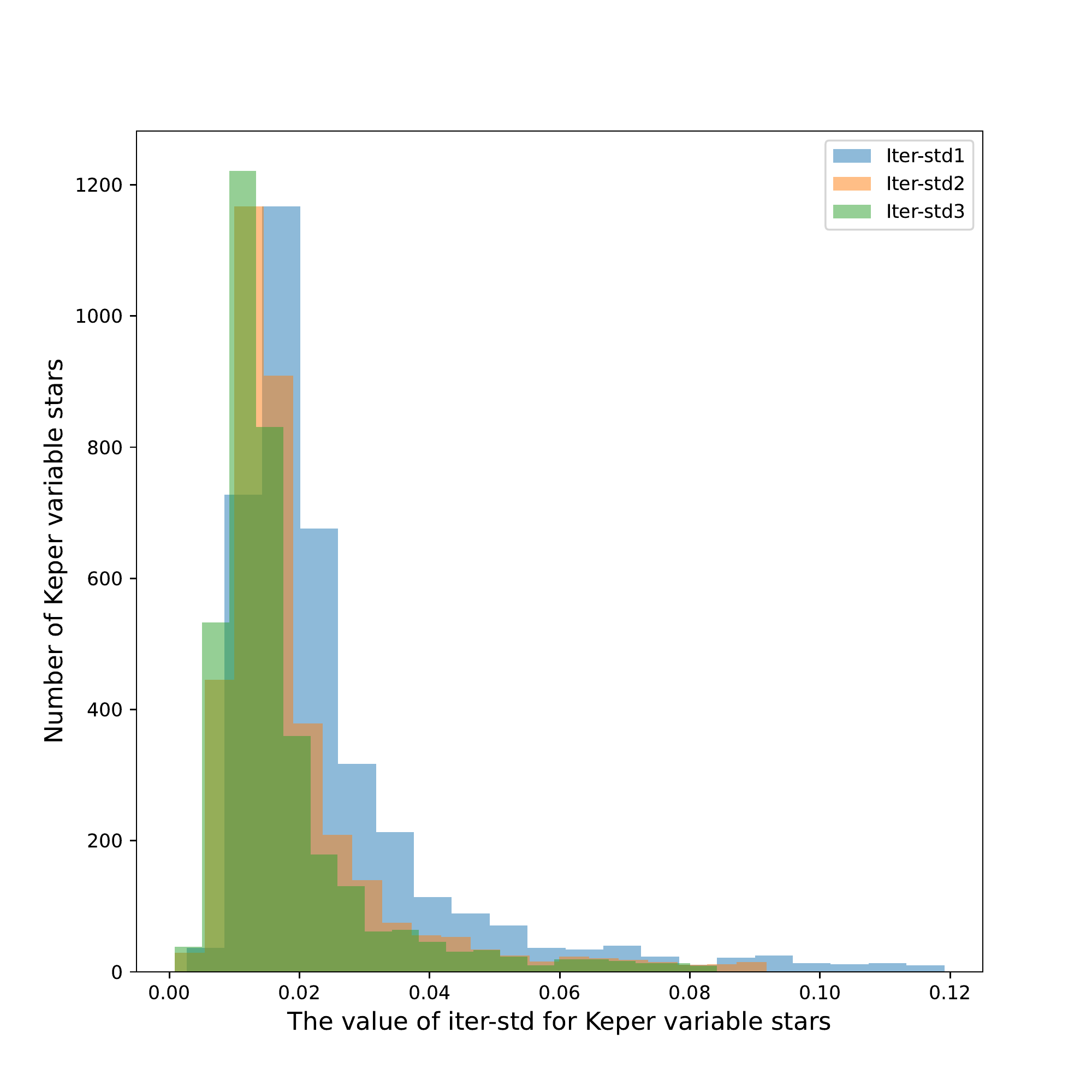} 
    \caption{The distribution of standard deviation under three iterations for Kepler variable sources.}
    \label{iter-std}
\end{figure}

Meanwhile, considering that the standard deviation may be affected by the anomalous data in the light curves, we removed the anomalous values larger than two times the standard deviation from the light curve data and then re-calculated the standard deviation. 
After three iterations, the standard deviation finally calculated was treated as Iter-std. The statistical histograms for different values of Iter-std within three iterations for each variable source are given in Figure~\ref{iter-std}. The distribution of the variable sources gradually concentrates toward the direction where the value of Iter-std becomes smaller with continuous iterations. Eventually, the position with the highest number of variable sources is stabilized at about 0.014.

\subsection{Statistics Modeling Based on Variability Parameters} \label{subsection:approach}

We take Q-value as an example and introduce the corresponding modeling process. Total 80$\%$ of the data in the sample set is used to perform statistical modeling, and the rest are used for model testing. 

1. Source labeling

The first step is labeling all sources based on whether the source is a variable source or not. The label of the source $L_{i}$ is given by: 

\begin{equation}
 L_{i}={
\begin{cases}
1& \text{ if } S_{i} \text{ is a variable source } \\ 
0& \text{ if } S_{i} \text{ is a non-variable source }
\end{cases}
, i \in [1,N]}
\end{equation}
where S is the sample set, $S_{i}$ is one source in the sample set. N is the number of sources in the sample set. 

2. Variability parameter calculation

For each source $S_{i}$, the corresponding variability parameter $V_{i}$ is given by:

\begin{equation}
V={\left\{ V_{i}  = 
     \frac{\left| {m_{max} - m_{min}} \right|} 
     {\sqrt{{\sigma_{max}^2 +\sigma_{min}^2}}}
     , i \in [1,N]
     \right\}
  }
\end{equation}
where V is the data set of the variability parameters. 

3. Variability probability calculation

At first, $V_{i}$ is sorted in ascending order, and $L_{i}$ is expanded sort based on $V_{i}$. Then the variability probabilities $P(V_{i})$ is calculated through the statistics of $L_{i}$ within an interval. This interval is that the 1$\%$ data of sample set adjacent to Vi. The $P(V_{i})$ is given by:

\begin{equation}
   P(V_{i})=\frac{\left| \left\{ L_{j}|\ L_{j}=1 \right\} \right|} 
     {\left| \left\{ L_{j}|\ L_{j}=1 \right\} \right| + \left| \left\{ L_{j}|\ L_{j}=0 \right\} \right|}
     ,  j \in [i-0.01*N, i+ 0.01*N], i \in [1,N]
\end{equation}

where, $P(V_{i})$ is the variability probabilities corresponding to $V_{i}$ . And $\left| \left\{ L_{j}|\ L_{j}=1 \right\} \right|$ is the cardinality of the set $\left\{ L_{j} |\ L_{j}=1 \right\}$ , $\left| \left\{ L_{j}|\ L_{j}=0 \right\} \right|$ is the cardinality of  the set $\left\{ L_{j} |\ L_{j}=0 \right\}$, respectively. 

\begin{figure}[ht]
    \centering
    \includegraphics[width=20cm,height =20cm]{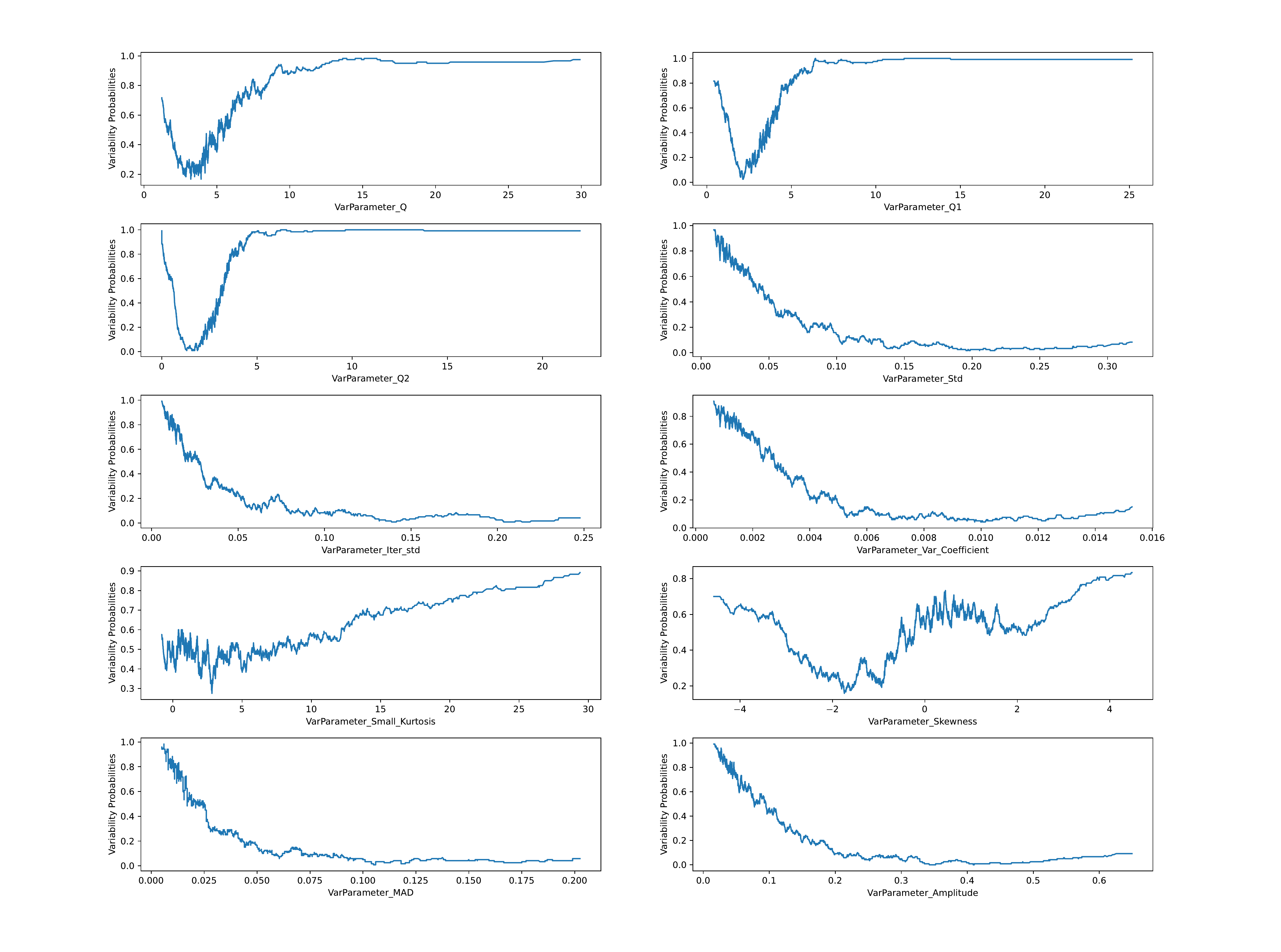}
    \caption{The variability probabilities at different variability parameters.}
    \label{fig-Precision}
\end{figure}

After the above three steps on the sample data set, we get the probabilities of the ten variability parameters at different values (see Fig.~\ref{fig-Precision}). These diagrams reflect the relationship between the variability parameters and the corresponding variability probabilities.

4. Variability parameters evaluation

\begin{figure}[h]
    \centering
    \includegraphics[width=18cm,height = 8cm]{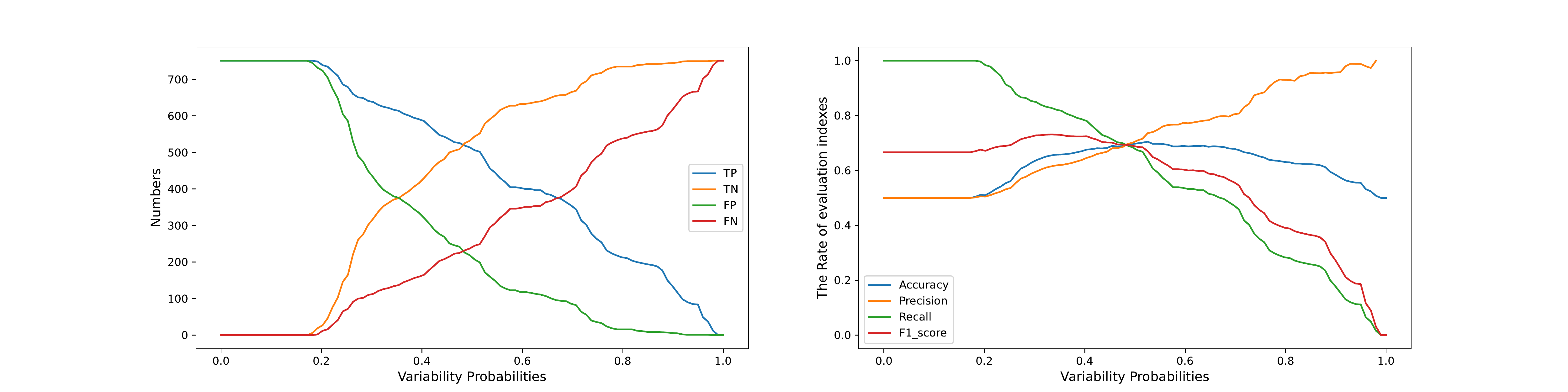} 
    \caption{The performance of variability parameter Q at different variability probabilities.}
    \label{fig-evluationindexes}
\end{figure}

We take the variability parameter Q as an example to demonstrate the evaluating procedures. The evaluation for other variability parameters is consistent and will not be repeated here.
Based on 20$\%$ of the data in the sample set, we counted the number of true positive (TP), true negative (TN), false positive (FP), and false-negative (FN) under 100 variability probabilities ranging from 0 to 1. The accuracy, precision, recall rate, and F1-score were also calculated (Fig.~\ref{fig-evluationindexes}). Meanwhile, the ROC curve area of variability parameter Q is 0.75.

\begin{figure}[h]
    \centering
    \includegraphics[width=14cm,height = 12cm]{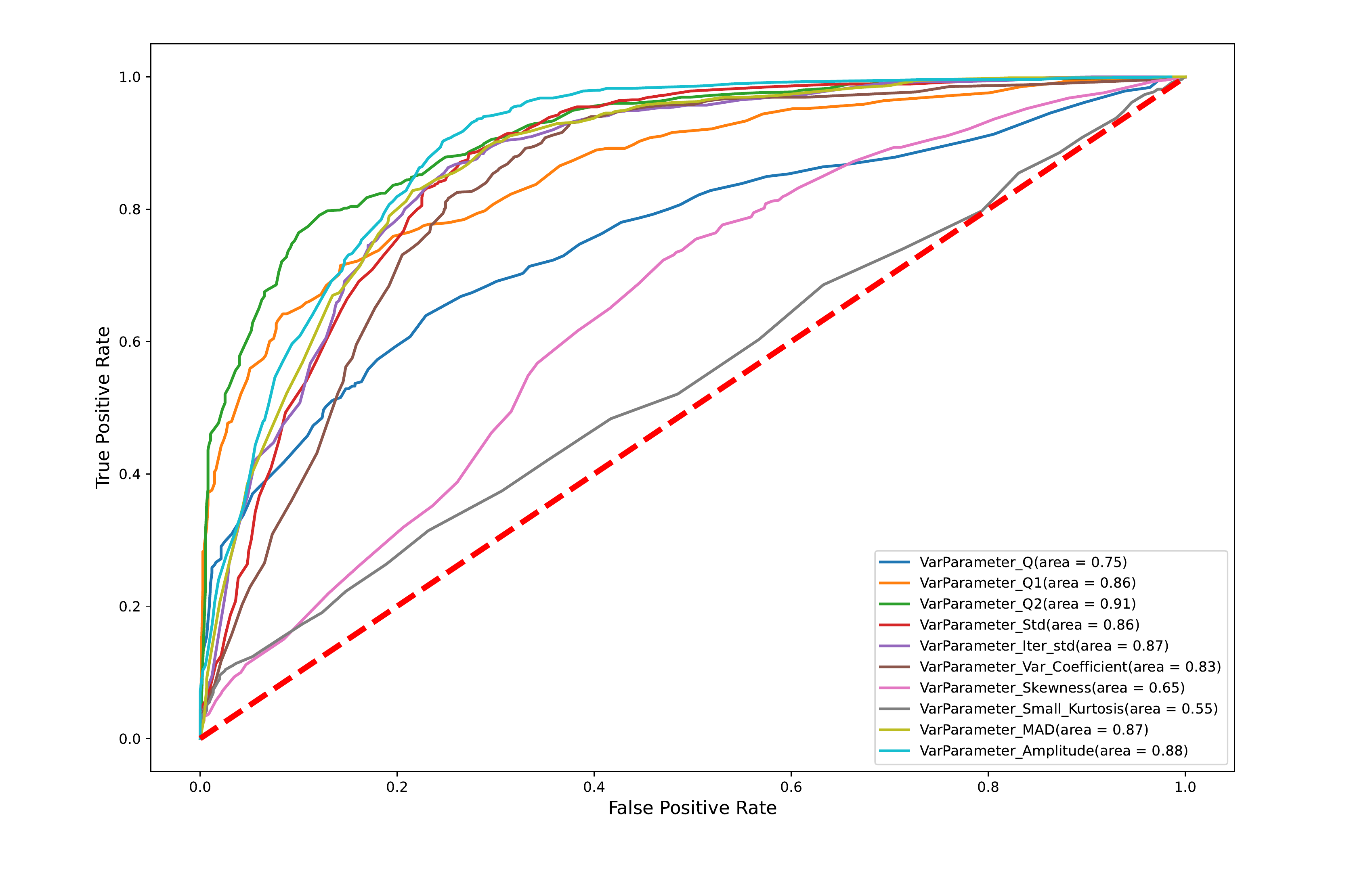} 
    \caption{The ROC curves of the ten variability parameters.}
    \label{fig-roc}
\end{figure}

The four credible evaluation indexes of the ten variability parameters models under the different probabilities are presented in Table ~\ref{evaluation indexes(0.68)} and Table ~\ref{evaluation indexes(0.95)}. Moreover, the areas of ROC curves are shown in Fig ~\ref{fig-roc}. According to the evaluation results, the Q2 parameters have better performance than the Q, Q1, Iter-std, Std, $C_{\nu}$, $\kappa$, $\gamma$, MAD and Amp parameters. Therefore, our final identification model for variable source candidates is given by parameters Q2. Of course, the other nine variability parameters can still be used as a reference.

\begin{table}[http]
\centering
\caption{The evaluation indexes of ten variability parameters (probability $\geq$ 0.68$\%$ be recognized as variable sources) \label{evaluation indexes(0.68)}}
\begin{tabular}{ccccc}
\hline\xrowht{20pt}
Parameters  & Accuracy    &  Recall   & Precision  &  F1 score \\
\hline
Q          & 0.69  & 0.46  & 0.84     &   0.55   \\
Q1         & 0.77  & 0.62  & 0.89  &  0.69    \\
Q2         & 0.83  & 0.72  & 0.92  &  0.77    \\
Std        & 0.77  & 0.71  & 0.80  &  0.74    \\
Iter-std   & 0.78  & 0.71  & 0.82  &  0.74   \\
$C_{\nu}$   & 0.76  & 0.73  & 0.78  &  0.75  \\
$\kappa$   & 0.54  & 0.10 & 0.80  &  0.17   \\
$\gamma$   & 0.53  & 0.09  & 0.73  &  0.16    \\
MAD        & 0.79  & 0.73  & 0.83  &  0.76    \\
Amp        & 0.76  & 0.73  & 0.78  &  0.75    \\
\hline
\end{tabular}
\end{table}

\begin{table}[http]
\centering
\caption{The evaluation indexes of ten variability  parameters(probability $\geq$ 0.95$\%$ be recognized as variable sources) \label{evaluation indexes(0.95)}}
\begin{tabular}{ccccc}
\hline\xrowht{20pt}
Parameters  & Accuracy    &  Recall   & Precision  &  F1 score \\
\hline
Q          & 0.53  & 0.07  & 0.98     &   0.12   \\
Q1         & 0.64  & 0.28  & 0.99  &  0.39    \\
Q2         & 0.69  & 0.38  & 0.99  &  0.49    \\
Std        & 0.52  & 0.04  & 1.00  &  0.07    \\
Iter-std   & 0.52  & 0.04  & 0.94  &  0.07   \\
$C_{\nu}$   & 0.50  & 0  & NaN  &  0   \\
$\kappa$   & 0.50  & 0 & NaN  &  0  \\
$\gamma$   & 0.50  & 0  & NaN  &  0    \\
MAD        & 0.51  & 0.03  & 0.88  &  0.06    \\
Amp        & 0.54  & 0.07  & 1.00  &  0.12    \\
\hline
\end{tabular}
\end{table}

\section{The Identification of LAMOST Sources} \label{sec:catalog}

The models of variable sources identification are constructed based on the variability parameters and statistical modeling. Therefore, in the follow-up LAMOST sources identification, we calculated the variability parameters from the light curves data. Moreover, we can obtain the final LAMOST variable source candidates based on the identification model, respectively.  

\subsection{Data Preparation} \label{sec:data}

We select the low-resolution data from LAMOST DR6 V1 with 9.91 million spectra, consisting of stars, galaxies, quasars, and other unknown objects. To ensure the quality of the spectral data, we limit the  SNRg $>$ 20 by referring to the related research of LAMOST data analysis (\cite{Liusnr20},\cite{Wangsnr30}). Finally, 4.68 million sources with high SNRg spectra and position information are extracted from LAMOST data and saved into an IPAC table.

The light curves data of these 4.68 million sources are matched from ZTF DR2 to identify variable sources. The ZTF DR2 contains light curves data acquired between 2018 March and 2019 June, covering a time span of around 470 days(\cite{masci2018zwicky},\cite{graham2019zwicky}). The ZTF DR2 includes more than 2 billion sources, about half of which have greater than 20 observations(\cite{bellm2019zwicky}). These light curves data can complement each other with the spectral data of the LAMOST, and the limiting magnitude and observational areas of the ZTF survey are close to those of the LAMOST survey. Referring to the pre-processing approach of the (\cite{ofek2020},\cite{Chenetal2020}), we merely use light curves of g and r band from the ZTF DR2 data set and matching with the IPAC table that has been generated previously. And the i band is not considered at all because there is no target source in ZTF DR2 with an detection times larger than 20.

At the same time, we attached two additional conditions to make the identification process of variable sources easier and quicker. 1) low-quality images and photometry data were excluded by adopting INFOBITS $<$ 33,554,432 and Catflags $\neq $ 32,768 (\cite{masci2018zwicky}). 2) We mainly selected sources in the ZTF DR2 data that were observed at least 50 times because the period’s false-alarm probability (FAP) is relatively high based on less than 50 detections. After restricting the above two conditions and eliminating duplicated sources, we finally obtained 2.66 million LAMOST data sources.

\subsection{Sources Identification and Variable Sources Catalog Generation} \label{sec:identify}

We perform variable source identification on LAMOST sources. After calculating the variability parameters obtained from light curves for each source, we determine whether the source is variable in the different variability parameters. Finally, we get the LAMOST variable source candidates based on the model of variability parameter Q2. As a result, all the catalogs of LAMOST variable source candidates in the different probabilities and conditions are shown in Table ~\ref{catalog}.

\begin{table}[http]
\centering
\caption{The catalogs of LAMOST variable source candidates($\alpha$ is the probability)\label{catalog}}
\begin{tabular}{ccc}
\toprule
\diagbox{Band}{Probability}  & $\alpha$ $\geq$ 95$\%$   &  $\alpha$ $\geq$ 68$\%$ \\
\midrule
g band    & 434,256   & 954,308 \\
r band    & 324,139    & 738,365  \\
\hline
Union of g and r band & 631,769 & 1,280,096 \\
\bottomrule
\end{tabular}
\end{table}

We take the catalog that contains 631,769 LAMOST variable source candidates as an example, and to list the detailed information are shown in Table ~\ref{catalogs}. 
This catalog contains the parameters provided by the LAMOST survey and the variability parameters obtained from the ZTF light curves data and corresponding probabilities. 

\begin{table}[h]
\centering
\caption{LAMOST variable source candidates catalog} \label{catalogs}
\setlength{\tabcolsep}{5.5pt}
\small
\begin{tabular}{cccccc}        
\toprule
R.A.(J2000)  &  Dec.(J2000) & ZTF$\_$oid(g/r)  & ... & Q2(g/r)  & P(Q2) (g/r) \\
\midrule

333.264311  &  1.835927   &     494101100004654/494201100007520   & ... &   40.00/45.04  &  0.99/0.99   \\
330.641667  &  1.239245   &     494102300003222/494202300005493   & ... &   9.51/7.52    &  0.99/0.98   \\
330.63137   &  1.835927   &     494102200004936/494202200009307   & ... &   23.54/22.27 &  0.99/0.99   \\
44.537916   &  0.952695   &     ---/453204400006321               & ... &   ---/4.95     &  ---/0.98    \\
10.08529    &   41.68036  &     695111200016244/695211200039474   & ... &   9.50/12.27   &  0.99/1.00   \\

10.214958   &  40.550835   &     695111300013377/695211300020792  & ... &   6.38/5.69   &  1.00/0.95    \\
81.4197442  &  29.6917296  &     ---/658201300022529              & ... &   ---/4.82    &  ---/0.98    \\
333.613593  &  29.758858   &     690101400006952/691204300010198  & ... &   5.14/4.98   &  0.98/0.98    \\
333.537014  &  30.355967   &     690101400008984/---              & ... &   5.64/---    &  0.95/---     \\
333.974481  &  29.998455   &     691104300005030/691204300016607  & ... &   9.25/11.73  &  0.99/1.00   \\

333.768941  &  29.761166  &     691104300007551/---               & ... &   4.71/---    &  0.97/---     \\
330.24329   &   30.66345  &     690102200008729/690202200013445   & ... &   14.87/13.69 &  0.99/1.00   \\
330.497222  &  31.192316  &     690102200001888/690202200002783   & ... &   39.74/31.76 &  0.99/0.99   \\
333.472737  &  32.11478   &     691208300004224/690105400002562   & ... &   8.65/12.08  &  0.99/1.00   \\
334.784969  &  31.591894  &     ---/691208400027388               & ... &   ---/5.01    &  ---/0.98    \\

330.698645  &  28.929285  &     ---/644215400001694               & ... &   ---/12.45    &  ---/1.00   \\
330.50609   &  29.093706  &     644115100020865/644215100014392   & ... &   6.69/10.41   &  0.99/1.00   \\
331.126246  &  28.929237  &     644114300001023/---               & ... &   5.85/---     &  0.96/---    \\
331.094036  &  29.315796  &     644114200006522/644214200010240   & ... &   6.04/4.76    &  0.98/0.97   \\
333.593922  &  32.639942  &     690105100006583/691208200010470   & ... &   6.66/6.11    &  0.99/0.99    \\

... &  ... & ...  & ...  & ... & ...  \\
... &  ... & ...  & ...  & ... & ...  \\
... &  ... & ...  & ...  & ... & ...  \\

\bottomrule
\end{tabular}
\begin{tablenotes}
\footnotesize
 \item[*] 20 lines are shown here for guidance regarding its form and content.
\end{tablenotes}
\end{table}

\section{Analysis and Evaluation for the Catalog} \label{sec:crossmatch}

In this section, we cross-identify variable source catalogs previously published to evaluate our catalog of 631,769 LAMOST variable source candidates. 

There are many catalogs of variable sources obtained from different survey projects, such as Kepler(\cite{Kirk2016EBcatalog}), GAIA(\cite{Clementinietal2019}, \cite{Mowlavietal2018}, \cite{Roelensetal2018}), LAMOST(\cite{Tianetal2020}) and ZTF(\cite{Chenetal2020},\cite{ofek2020}). These catalogs contained variable sources of different types, such as period variable sources, eclipsing binaries, RV variable sources, and so on. Thus, we use cross-identification to search for common sources between published variable source catalogs and LAMOST variable sources identified in the study and verify their correctness.

\subsection{Cross-match with GAIA Variable Sources} \label{sec:cross-match GAIA}

The Gaia delivers nearly simultaneous measurements in the three observational domains on which most stellar astronomical studies are based: astrometry, photometry, and spectroscopy(\cite{brown2016gaia}). The Gaia data releases provide accurate astrometric measurements for an unprecedented number of objects. In particular, trigonometric parallaxes carry invaluable information since they are required to infer stellar luminosities, which form the basis of understanding much of stellar astrophysics. 

\begin{figure}[htb]
    \centering
    \includegraphics[width= 13cm,height = 9cm]{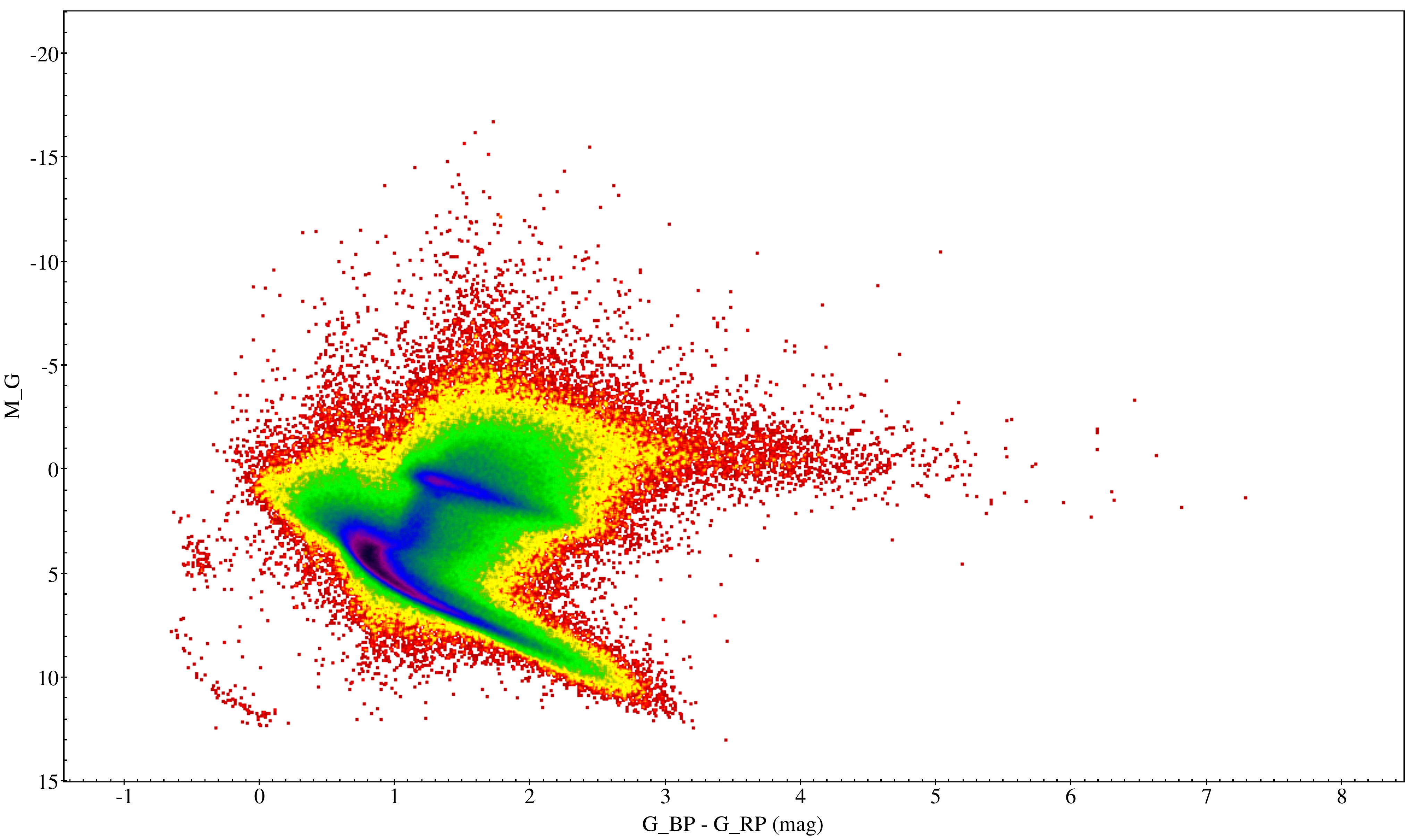} 
    \caption{The CaMD of the LAMOST variable source candidates. The G\_BP - G\_RP is color of these candidates in the G band of Gaia DR2. The absolute Gaia magnitude in the G band for individual sources are estimated through by M\_G  = G + 5 + 5*$\log_{10}(\varpi/1000)$ , with G is magnitude in the G band,  $\varpi$ is the parallax in milliarcseconds(\cite{GAIADR22018A&A})}
    \label{CaMDs50}
\end{figure}

Based on these properties, the GAIA survey data can be complemented and verified with other telescope projects. The catalogs of LAMOST variable source candidates obtained in the study are also cross-match with the Gaia DR2 catalog in the Tool for OPerations on Catalogues And Tables(TOPCAT) software(\cite{2005ASPC34729T}). The location distribution of LAMOST variable source candidates is obtained through Colour-absolute Magnitude Diagram(CaMD) as well (see Fig.\ref{CaMDs50}). 

It can be seen from Fig.\ref{CaMDs50} that the LAMOST variable source candidates in our catalog are concentrated in the main sequence belt, followed by the giant star branch. In contrast, the white dwarf star sequence has fewer sources, which is consistent with the overall distribution of stars.

Gaia data include the Cepheid(Cep), RR Lyrae(RRL), long-period variable(LPV), short-period variable(SPV) (\cite{Clementinietal2019}, \cite{Mowlavietal2018}, \cite{Roelensetal2018}). There are 2106 common sources between the catalog of GAIA variable sources catalogs and LAMOST identification targets, and these sources are recognized as variable sources in our catalogs. The detection rate is 100$\%$ through our variability parameter models. Among them, these sources is contain the 370 LPVs, 60 Ceps, 1667 RRLs and 8 SPVs.

\subsection{Cross-match with Catalog of \cite{Chenetal2020}}\label{subsection:matchZTFDR2}

ZTF DR2 represents a highly suitable database for the detection and exploration of new variable source candidates. The catalog of 781,602 periodic variables was published by \cite{Chenetal2020}. Comparison with previously published catalogs shows that 621,702 objects (79.5$\%$) are newly discovered or newly classified, including $\sim$ 700 Cepheids, $\sim$ 5000 RR Lyrae stars, $\sim$ 15,000 $\delta$ Scuti variables, $\sim$ 350,000 eclipsing binaries, $\sim$ 100,000 long-period variables, and about 150,000 rotational variables. 

\begin{figure}[http]
    \centering
    \includegraphics[width=8.9cm,height = 8cm]{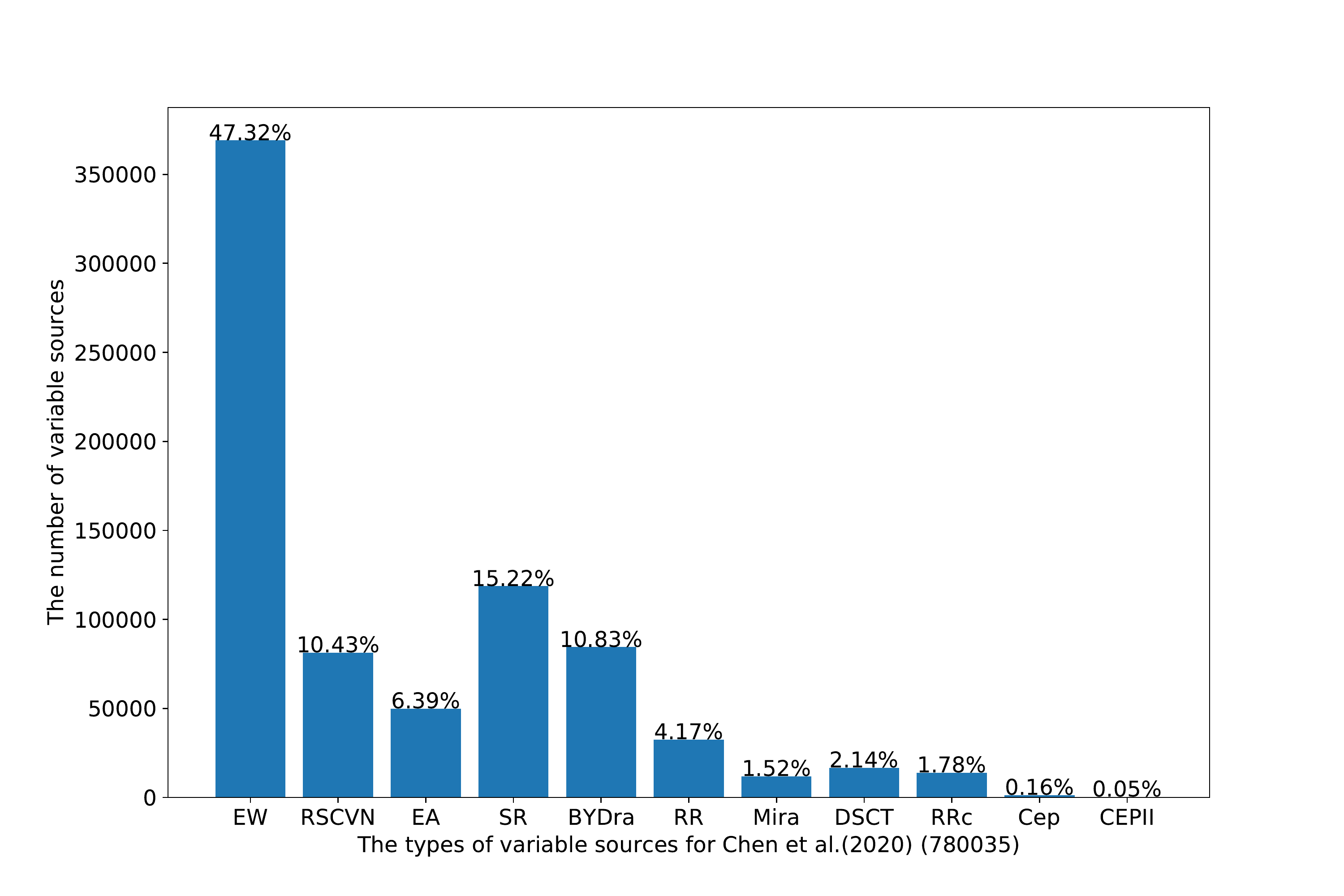} 
    \includegraphics[width=8.9cm,height = 8cm]{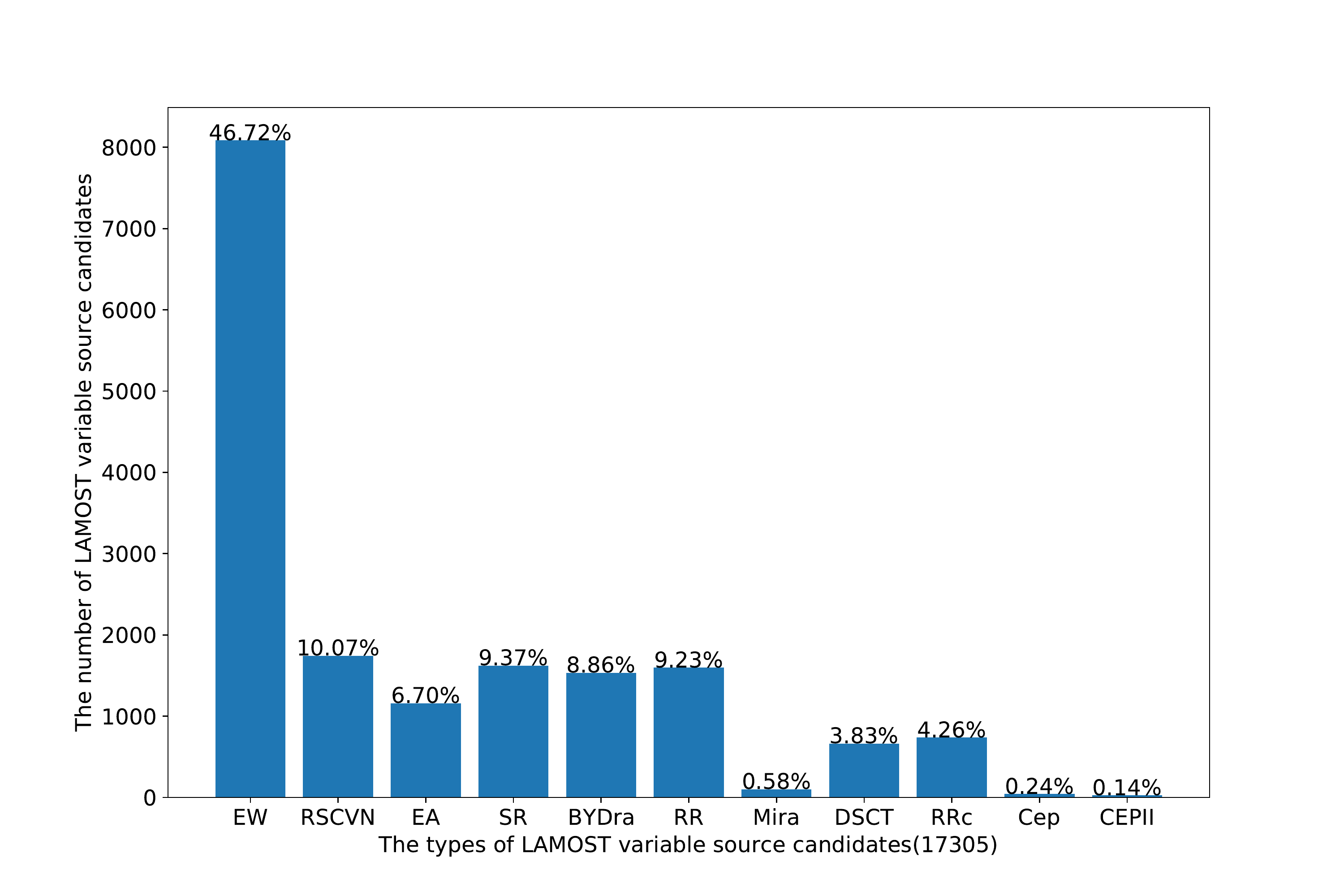} 
    \caption{The types distribution of variable sources (The left is  \cite{Chenetal2020}, and the right is Our catalog cross-match with \cite{Chenetal2020}).}
    \label{variable source types}
\end{figure}

However, we only selected the sources crossed between ZTF DR2 and LAMOST DR6 V1 and imposed a strict limit on the number of observations. Therefore, the data set used in our work is only a tiny fraction of the complete ZTF DR2. Nonetheless, there are 17,711 common sources between the catalog of ZTF DR2 periodic variable sources(PVS) and LAMOST identification targets. Among them, 17,305 objects are recognized as PVS under a probability greater than 95$\%$. Based on that, the detection rate of PVS is about 98$\%$ through our variability parameter models. The distribution of the variable sources types in our catalog is shown in Fig.~\ref{variable source types} through the cross match \cite{Chenetal2020}. It is shown that our variability parameter model is valid for the most common types of variable sources.

\subsection{Cross-match with Catalog of \cite{ofek2020}}\label{subsection:matchZTFDR1}

\begin{figure}[h]
    \centering
    \includegraphics[width=13cm,height = 9cm]{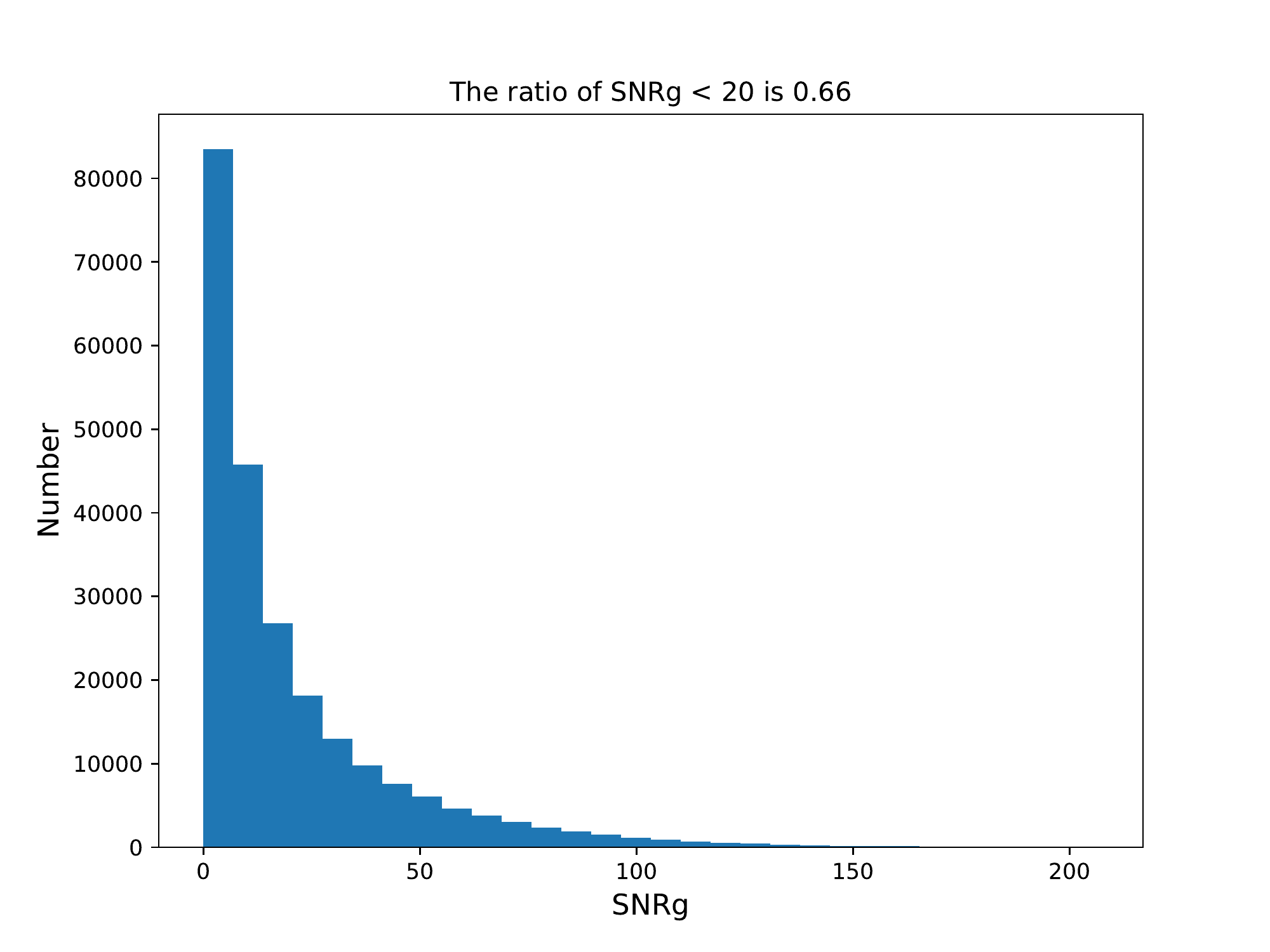} 
    \caption{The SNRg distribution of LAMOST variable sources presented by \cite{ofek2020}}
    \label{SNRg_distribution}
\end{figure}

The catalog presented by \cite{ofek2020} includes 10 million variable sources, the largest published catalog of ZTF variable sources. However, as mentioned in the previous section, only 216,007 sources can be cross-matched with LAMOST. The distribution diagram of SNRg of these sources is shown in Fig.~\ref{SNRg_distribution}. After removing data with SNR less than 20 and data points in the light curves less than 50, there are 71,454 common sources. Our model successfully identified 46,073 variable sources with a success rate of 64\%. We also conducted a cross-match of the three variable catalogs. There are 7534 common sources among \cite{ofek2020}, \cite{Chenetal2020} and our LAMOST identification targets.  7383 sources are included in our catalog.

We further investigated the reasons that theses sources were not identified. We manually analyzed the light curves of a random sample of these sources. We note that the magnitude error of each photometric observation is given in the ZTF catalog. The magnitude variations of these unidentified sources are basically within the observed magnitude errors. Therefore, we realize that the difference between our results and \cite{ofek2020} is due to the definition of variable sources. The variable sources identified in this study must ensure two observations with magnitude variations exceeding a factor of 2 magnitude error. While \cite{ofek2020} selected variable candidates mainly give specific thresholds under the robust std or classical periodogram.

To validate this analysis, we directly examined the variable source data given by Ofek using our given variability parameter ($P(Q2)\geq68\%$). We randomly selected two-variable source sets from \citep{ofek2020}. Table 2 shows that our model successfully identified 91\% objects, proving that the Q2 is robust for variables identification.

\begin{table}[http]
\centering
\caption{The performance of variability parameter Q2 for variable sources catalog of \cite{ofek2020} \label{modeltest_ofek}}
\begin{tabular}{ccc}
\toprule
\makecell{Number of variable sources that randomly\\ selected from \cite{ofek2020}}  & \makecell{Number of variable sources  \\ identified by our model} &   Rate \\
\midrule
10,029    & 9,087   & 0.91 \\
474,186   & 431,428    &  0.91  \\
\bottomrule
\end{tabular}
\end{table}

\subsection{Cross-match with Catalog of \cite{Kirk2016EBcatalog}}\label{subsection:matchKEPLER}
The Kepler Mission provided nearly continuous monitoring of $~$200,000 objects with unprecedented photometric precision. The catalog of eclipsing binary systems within the 105 deg$^\circ$ Kepler field of view was published by \cite{Kirk2016EBcatalog}. This catalog lists the KIC, ephemeris, morphology, principle parameters, and so on. There are 859 common sources between the Kepler Eclipsing Binaries(EBs) and the variable catalog proposed in the study. Among them, 662 sources are identified as eclipsing binaries under a probability greater than 95$\%$. The detection rate of EBs is about 77$\%$ through our statistical model.

\subsection{Cross-match with Catalog of \cite{Tianetal2020}} \label{subsection:matchLAMOST}
LAMOST spectroscopic survey has provided $\sim$ 4.7 million unique sources were targeted and $\sim$ 1 million stars observed repeatedly. The probabilities of stars being RV variables are estimated by comparing the observed radial velocity variations (RVVs) with simulated ones. The catalog of RVVs based on the LAMOST survey has been published by \cite{Tianetal2020}. This catalog collects 80702 variable sources, including 77 $\%$ binary systems and 7$\%$ pulsating stars as well as 16$\%$ pollution by single stars. There are 24,092 common sources between the LAMOST RVVs catalog and LAMOST identification targets. Among them, 12,141 objects are recognized as RVVs under a probability greater than 95$\%$. The total detection rate of RVVs is about 50$\%$ through our variability parameter models.

At the same time, we also performed an additional cross-checked with other catalogs. There are 12,141 common sources between our catalog with \cite{Tianetal2020} with a rate of 1.9\%, 1500 common variable sources between the \cite{Chenetal2020} and \cite{Tianetal2020} with a rate of 0.19\%, and 686 common variable sources between \cite{ofek2020} and \cite{Tianetal2020} with a rate of 0.0064\%.

\subsection{Performance of the Catalog} \label{subsection:Performance}
A summary of the numbers of common sources between the published catalogs and LAMOST identification targets is listed in Table ~\ref{performance}. Note that some variable sources are identified repeatedly in different published variable sources catalogs. There are 123,756 common sources between our sources to be identified and the published catalogs such as GAIA DR2 long-period variable sources, LAMOST RV variable sources, KEPLER eclipsing binaries, and ZTF DR2 period variable sources. Total 85,669 sources are detected as variable sources in our catalog under a probability greater than 95$\%$, and are classified as different types. The detection rate of our catalog is 69$\%$ for the variable sources published in the referred catalogs.

\begin{table}[htp]
\centering
\caption{The cross-identification between our catalog and published catalogs\label{performance}}
 \setlength{\tabcolsep}{0.005pt}{
\begin{tabular}{cccc}
\toprule
Published Catalogs  & \makecell{Common sources with \\LAMOST identification targets} & \makecell{Identified by our model\\($\alpha$ $\geq$ 0.95)}  &  \makecell{Identified rate}\\
\midrule
\cite{Clementinietal2019},etc.               &  2106    &  2105      & 1.00   \\
\cite{Chenetal2020}                          &  17711   &  17305     & 0.98   \\
\cite{ofek2020}                              & 71454   & 46073     & 0.64   \\
\cite{Chenetal2020} $\ast$ \cite{ofek2020}   & 7534    & 7383      &0.98    \\
\cite{Kirk2016EBcatalog}                     &  859    &  662      & 0.77   \\
\cite{Tianetal2020}                          & 24092   &  12141     & 0.50  \\
\hline
Total                               & 123756   &  85669    &0.69   \\
\bottomrule
\end{tabular}}
\begin{tablenotes}
\footnotesize
 \item[*]  $\ast$ is indicate the cross-match between two variable source catalogs, and $\alpha$ is the probability.
\end{tablenotes}
\end{table} 

\section{Discussion} \label{sec:Discussion}

\subsection{Data Sets with Different Bias}\label{samplesets}

The data sets containing the variable sources and standard stars are from different catalogs in the statistical modeling, which may have some potential biases due to the number of observations, types of source, distribution of magnitude, and so on. However, some minor biases exist from the same data set due to the influence of observation conditions, i.e., weather conditions and poor seeing. Therefore, constructing a completely unbiased data set is very difficult. The study obtained the corresponding light curves of these sources collected from the ZTF DR2 by cross-matching other catalogs, but we only rely on the ZTF light curve data.

In addition, our non-variable sources in the study were derived from the SDSS catalog of standard stars. Some new research work suggests that some standard stars from different survey projects may be less standard, which may affect the final results of this study.
For this reason, in the statistical modeling, we only randomly selected standard stars with the same number of variable and non-variable sources. In addition, we performed some experiments to replace the samples in the original data set with randomly selected data (1/2, 1/3, and 1/4) from the standard star catalog. We find a slight change in the evaluation index, but the variability parameter Q2 still performs the best, indicating that the results of this study are plausible.

\subsection{Variable Source Identification Method}\label{Method limitations}

The identification method of variable sources is a very classical problem, and there has been a lot of preliminary research work. These methods are generally based on multiple photometric data, calculating the magnitude variance to determine whether it is a variable source, or using period analysis to determine whether it is a periodic variable source. The variable source identification method presented in the study draws fully on these methods. In the statistical modeling, we extract the corresponding targets in the ZTF by using the variable sources confirmed in the Kepler catalog and statistically modeling the light curves data of these targets. Compared with the classical method, the calculation is fast and straightforward, well-suited for variable source identification in large-scale catalogs. Meanwhile, the experimental results show that the obtained accuracy is at least comparable to the conventional method. In addition, the statistical modeling method in this paper is a general identification method for different types of variable sources, which facilitates the classification of variable sources.

Of course, there are some limitations to this statistical modeling approach. For example, the number of observations of a source can directly affect the correctness of the identification. At the same time, a large accidental error during one photometry may lead to a final misjudgment.
In addition, our method does not sensitive for some special variable sources(outbursting stars, cataclysmic variables and so on). For these types of variable sources, specific methods may be needed for identification, and we will consider this as a future work.

\subsection{The Correlation Analysis of Variability Parameters} \label{sec:Correlation}

\begin{figure}[h]
    \centering
    \includegraphics[width=15cm,height = 12cm]{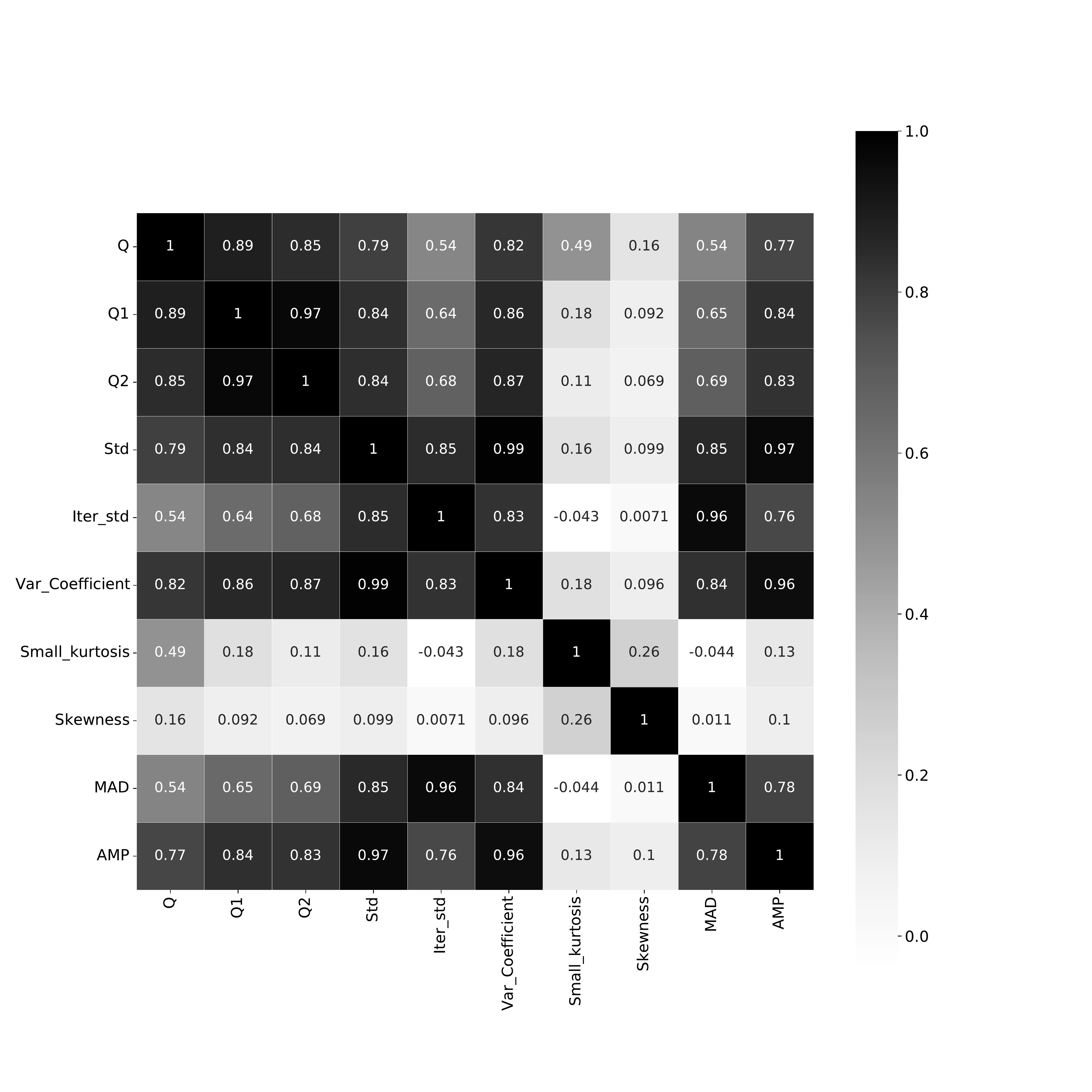} 
    \includegraphics[width=15cm,height = 12 cm]{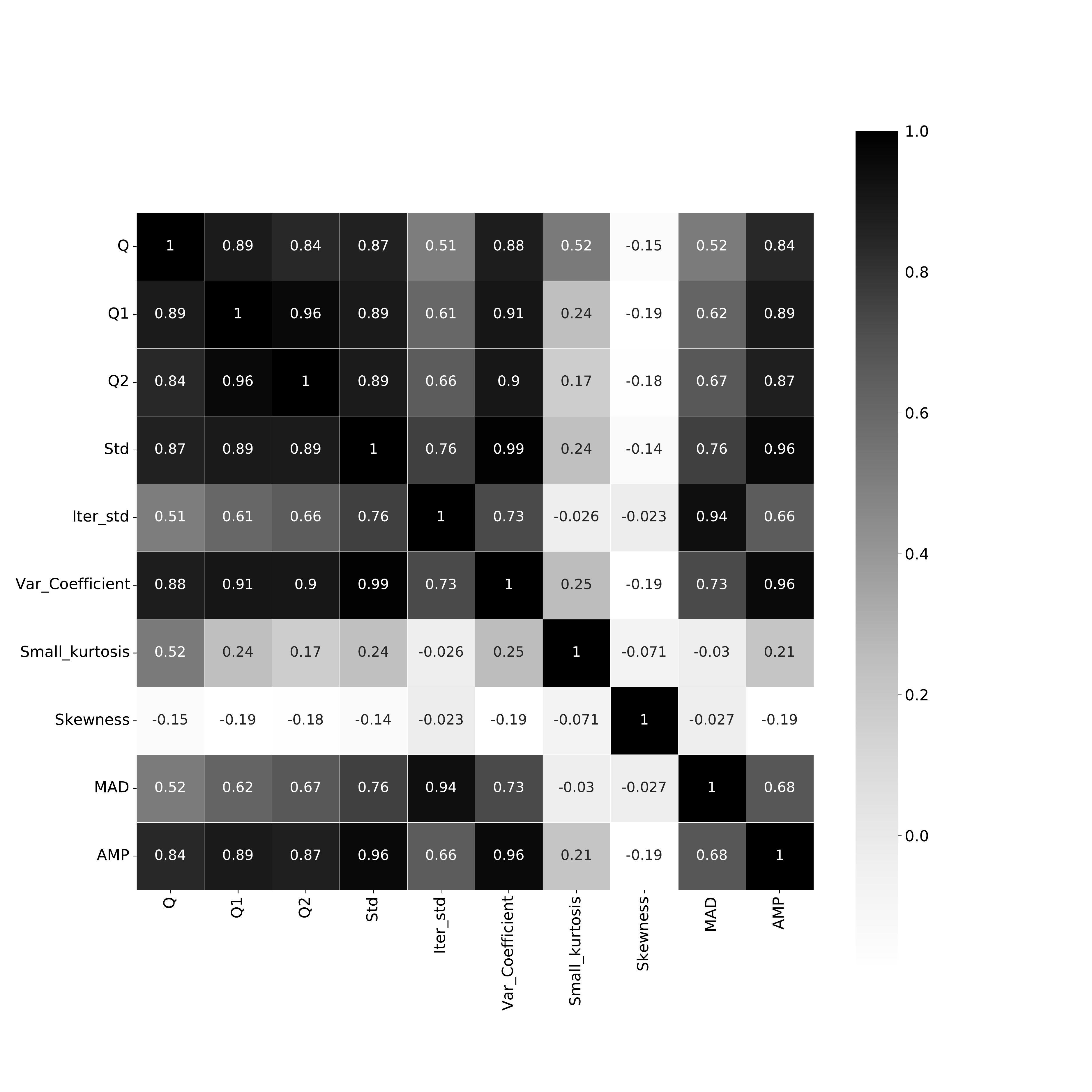} 
    \caption{The Correlation matrix ten variability parameters for variable source candidates in two band of ZTF (Top: g band, Bottom: r band.)}
    \label{corr_analysis}
\end{figure}

We performed the correlation analysis on the ten variability parameters using for modeling, and the corresponding correlation coefficient matrix is shown in (Fig.\ref{corr_analysis}). We can know that linear relationships between all features to some extent, and most of them are positively correlated. Of course, these variability parameters still have other functional relationships, such as square relationships, logarithmic relationships, and so on.

\subsection{The Limitations of Variability Parameter Models} \label{subsec:Model  Limit}
In the process of variable source identification, ten variability parameters (i.e., Q, Q1, Q2, Std, Iter-std, $C_{\nu}$, $\kappa$, $\gamma$, MAD and Amp) are calculated based on the light curves data obtained from ZTF DR2. These parameters are intrinsic statistical properties relating to the scale(Std, Iter-std), morphology($\kappa$, $\gamma$, Amp), or other properties. These parameters are highly explainable and robust against bias (\cite{cabraletal2018}). However, they may also have varying utility. For example, the lack of some statistical attributes may affect the results of variable source identification. For example, the time statistical properties(Period) can reflect more information in the time series data. In future work, we must consider introducing more parameters that contain more helpful information to identify variable sources.

\section{Conclusion} \label{sec:Conclusion}
The LAMOST survey has accumulated massive amounts of spectral data, but the research on variable sources is limited to a certain extent due to the lack of photometric information. We combined the light curves data obtained from ZTF time-domain survey to identify variable source candidates in this work. A catalog of 631,769 LAMOST variable source candidates is constructed with a probability greater than 95$\%$, and this process is based on the variability parameters and statistical modeling for light curves data. This catalog is a robust database of variable source candidates. Moreover, the methods in this work are not only for the identification of periodic variables but also for general variables, including transient sources. They will be beneficial in a few future time-domain survey projects, such as WFST and MEPHESTO, in construction in China. 

Meanwhile, the cross-identification and classification are carried out by matching with the published catalogs.  Total 85,669 variable sources(the probability greater than 95$\%$) in our catalog are identified and classified, containing the cepheid, RR lyrae, Eclipsing binaries, long-period, short-period, RV variable sources, and so on. Although recognized as variable sources, most of our catalog objects are not classified based on data-match with other catalogs.

In further work, we will use spectral data and photometric information from LAMOST and other time-domain surveys to classify the catalog of variable sources as a follow-up to this work. Meanwhile, We will use machine learning methods for variable sources classification based on the calculated variability parameter of light curves data. Besides, LAMOST has officially entered the mid-resolution surveys and accumulating massive spectral data. At the same time, ZTF DRs will cover both increased timespans and large numbers of exposures. These aspects will help improve the determination of the long-period variables and increase the completeness of the catalog. In addition to the main types of variables discussed here, there are many cataclysmic variables, low-amplitude pulsating stars, binaries with compact objects, and stars with exoplanets. We will identify these variables to construct a large sample for further study based on future ZTF DRs and LAMOST DRs.

\normalem 
\begin{acknowledgements} 

We thank the anonymous referee for valuable and helpful comments and suggestions. This work is supported by the National SKA Program of China No 2020SKA0110300, the Joint Research Fund in Astronomy (U1831204, U1931141) under cooperative agreement between the National Natural Science Foundation of China (NSFC) and the Chinese Academy of Sciences (CAS), National Science Foundation for Young Scholars (11903009). Funds for International Cooperation and Exchange of the National Natural Science Foundation of China (11961141001). Fundamental and Application Research Project of Guangzhou(202102020677). The Innovation Research for the Postgraduates of Guangzhou University (2020GDJC-D20). This work is also supported by Astronomical Big Data Joint Research Center, co-founded by National Astronomical Observatories, Chinese Academy of Sciences and Alibaba Cloud.

\end{acknowledgements}

\bibliography{xtt}{}
\bibliographystyle{aasjournal}
\end{document}